\documentclass[%
 reprint,
superscriptaddress,
nofootinbib,
 amsmath,amssymb,
prb,
floatfix,
]{revtex4-1}
\usepackage[utf8]{inputenc}
\usepackage[T1]{fontenc} 

\usepackage{graphicx}% Include figure files
\usepackage{dcolumn}% Align table columns on decimal point
\usepackage{bm}% bold math
\usepackage{braket}
\usepackage{color}
\usepackage{multirow}

\usepackage{epstopdf}

\renewcommand{\t}{\text}

\renewcommand{\vec}[1]{\textbf{#1}}

\newcommand{\MoS}{MoS$_2$}

\begin{document}

\preprint{APS/123-QED}

\title{
	Non-invasive control of excitons in two-dimensional materials}

\author{C. Steinke}
	\email{csteinke@itp.uni-bremen.de}
	\affiliation{Institut f{\"u}r Theoretische Physik, Universit{\"a}t Bremen, Otto-Hahn-Allee 1, 28359 Bremen, Germany}
	\affiliation{Bremen Center for Computational Materials Science, Universit{\"a}t Bremen, Am Fallturm 1a, 28359 Bremen, Germany}
 
\author{D. Mourad}%
	\affiliation{Institut f{\"u}r Theoretische Physik, Universit{\"a}t Bremen, Otto-Hahn-Allee 1, 28359 Bremen, Germany}
	\affiliation{Bremen Center for Computational Materials Science, Universit{\"a}t Bremen, Am Fallturm 1a, 28359 Bremen, Germany}
	\affiliation{Present adress: Thermo Fisher Scientific (Bremen), Hanna-Kunath-Strasse 11, 28199 Bremen, Germany}
	
\author{M. R\"osner}
	\affiliation{Institut f{\"u}r Theoretische Physik, Universit{\"a}t Bremen, Otto-Hahn-Allee 1, 28359 Bremen, Germany}
	\affiliation{Bremen Center for Computational Materials Science, Universit{\"a}t Bremen, Am Fallturm 1a, 28359 Bremen, Germany}
	\affiliation{Department of Physics and Astronomy, University of Southern California, Los Angeles, California 90089-0484, USA}
	
\author{M. Lorke}
\author{C. Gies}
\author{F.Jahnke}
\author{G.Czycholl}
\affiliation{Institut f{\"u}r Theoretische Physik, Universit{\"a}t Bremen, Otto-Hahn-Allee 1, 28359 Bremen, Germany}
\author{T. O. Wehling}
	\affiliation{Institut f{\"u}r Theoretische Physik, Universit{\"a}t Bremen, Otto-Hahn-Allee 1, 28359 Bremen, Germany}
	\affiliation{Bremen Center for Computational Materials Science, Universit{\"a}t Bremen, Am Fallturm 1a, 28359 Bremen, Germany}

\date{\today}% It is always \today, today,

\begin{abstract}
We investigate how external screening shapes excitons in two-dimensional (2d) semiconductors embedded in laterally structured dielectric environments. An atomic scale view of these elementary excitations is developed using models which apply to a variety of materials including transition metal dichalcogenides (TMDCs). We find that structured dielectrics imprint a peculiar potential energy landscape on excitons in these systems: While the ground-state exciton is least influenced, higher excitations are attracted towards regions with high dielectric constant of the environment. This landscape is ``inverted'' in the sense that low energy excitons are less strongly affected than their higher energy counterparts. Corresponding energy variations emerge on length scales of the order of a few unit cells. This opens the prospect of trapping and guiding of higher excitons by means of tailor-made dielectric substrates on ultimately small spatial scales. 
\end{abstract}

%Use showkeys class option if keyword
\maketitle

%\tableofcontents

%-------------------------------------------------------------------
%%%%%%%%%%%%%%%%%%%%%%%%%%%%%INTRODUCTION%%%%%%%%%%%%%%%%%%%%%%%%%%
%-------------------------------------------------------------------

\section{Introduction}
\begin{figure*}[htp]
		\includegraphics{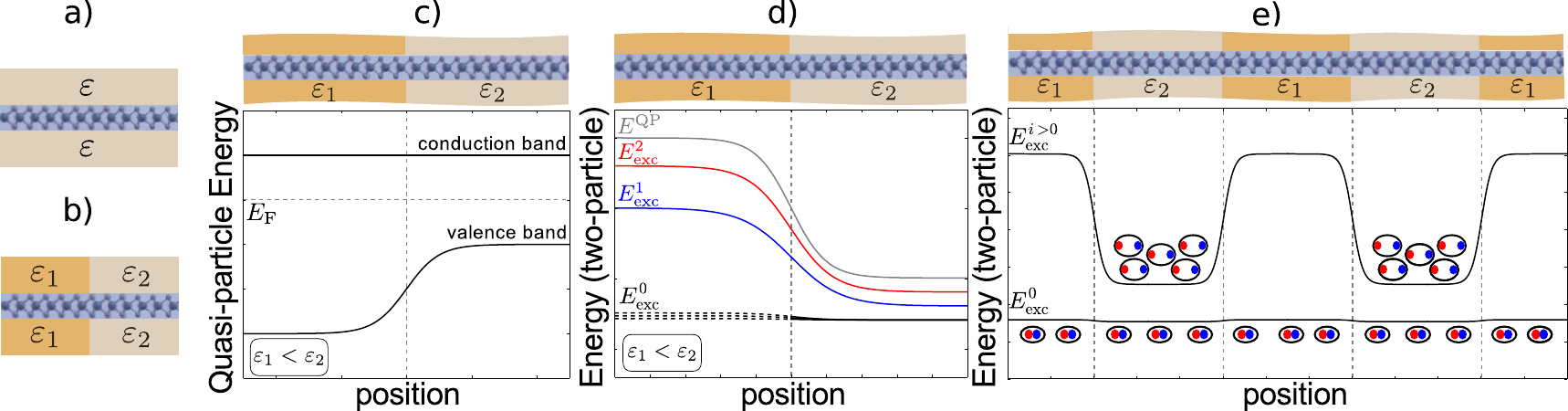}
		\caption{\label{fig:environment}(Color online) A monolayer of a 2d semiconductor (blue) embedded in different dielectric environments: a) A homogeneous environment with macroscopic dielectric constant $\varepsilon$. b) A laterally structured dielectric environment with different dielectric constants $\varepsilon_1$ and $\varepsilon_2$. 
		The heterogeneous environment sketched in b) imprints a potential energy landscape on electronic quasi-particles (panel c) and electron-hole excitations / excitons (panel d). Higher energy excitons are attracted towards regions with higher dielectric constant. The lowest-energy excitons are affected more weakly and can either attracted towards or expelled from regions with higher dielectric constant. Therefore, nanostructured dielectric substrates allow for the realization of potential energy landscape for excitons as shown in panel e). Under optically pumping, these can yield "inverted" regions where higher energy excitonic states have a higher occupation than the ones at lower energies.
		}
	\end{figure*}
	
Coulomb interaction causes pronounced correlation phenomena such as superconductivity\cite{somoano_alkaline_1975,staley_electric_2009,shi_superconductivity_2015,jo_electrostatically_2015,schonhoff_interplay_2016}, charge-density waves\cite{xi_strongly_2015,wilson_charge-density_1974,castro_neto_charge_2001,rossnagel_origin_2011}, magnetism\cite{wei_electronic_2015,tongay_magnetic_2012,peres_coulomb_2005,mcguire_coupling_2015} and strong excitonic effects\cite{ceballos_exciton_2016,qiu_optical_2013,ramasubramaniam_large_2012,steinhoff_influence_2014,ugeda_giant_2014,komsa_effects_2012,chernikov_exciton_2014} in two-dimensional (2d) materials. Monolayers of these materials realize atomically thin electronic systems, where the Coulomb interaction is strongly dependent on the dielectric environment as for instance the substrate of the material \cite{ugeda_giant_2014,komsa_effects_2012,hill_observation_2015,chernikov_exciton_2014,qiu_optical_2013,huser_how_2013,rosner_wannier_2015,park_first-principles_2009,lin_dielectric_2014,chen_dielectric_2009,park_angle-resolved_2009,ryou_monolayer_2016,bradley_probing_2015,zhang_bandgap_2016,stier_probing_2016}. This leads to the exciting opportunity to control interaction driven material properties externally and non-invasively via screening of the substrate or some adsorbates: for instance, laterally structured substrates (see Fig.~\ref{fig:environment}) can be used to create "junctions" with band gap modulations on the scale of several $100$~meV within one and the same material.~\cite{rosner_two-dimensional_2016,raja_coulomb_2017} For independently moving electrons or holes, indeed very sharp potential energy modulations on the scale of a few nm or even less are possible.~\cite{rosner_two-dimensional_2016} However, under optical excitation electrons and holes are known to form strongly bound excitons instead of moving independently in 2d materials. It is thus unclear which optoelectronic material functionalities can be imprinted externally. To change this situation, it is central to understand how and at which length scales excitons are influenced by laterally structured dielectric environments.

In this paper, we investigate how excitons respond to the tuning of the Coulomb interaction in 2d materials. To this end, we consider two tight-binding model systems, which emulate single semiconducting layers embedded in a dielectric environment. 
We study two different kinds of environments as depicted in Fig.~\ref{fig:environment}: We first analyze the influence of different homogeneous dielectric environments, characterized by the  macroscopic dielectric constant $\varepsilon$ of the substrate, on our monolayer and show that the size of the exciton Bohr radius $a$ determines how two-particle excitations react to changes in the materials dielectric environment.

Then, we investigate how a spatially structured dielectric environment influences the energy spectrum of an excited system. We show that any exciton which has a considerable spread is strongly affected by the environment, giving rise to the prospect of atomic scale trapping and guiding of higher excited excitons by means of laterally structured dielectric substrates. The only exception holds for the energy of strongly localized excitons, which reveal a non-monotonous behavior leading to interesting possibilities for nanoscale engineering including the creation of novel quantum confined materials. These might serve as building blocks for nanoscale lasers and could help realizing exotic states of excitonic matter (cf. Fig.~\ref{fig:environment}e for a visual account of a possible potential energy landscape).

%-------------------------------------------------------------------  
%%%%%%%%%%%%%%%%%%%%%%%%%%%%%%%METHODS%%%%%%%%%%%%%%%%%%%%%%%%%%%%%
%-------------------------------------------------------------------
 
\section{Modelling spatially resolved optical properties} \label{sec:Methods}
	To investigate the influence of dielectric environments on excitonic effects in 2d materials we diagonalize a many-body Hamiltonian in the electron-hole picture. We consider a two-band model and the excitation of one electron from the valence to the conduction band. The interaction-free ground-state is described in the tight-binding approximation with a basis set $\lbrace{\ket{\vec{R}}\rbrace}$, where $\vec{R}$ is labeling the lattice site on which the orbital is predominantly localized. To account for the electron-electron Coulomb interaction effects on the ground state and on single-particle excitations, we use the Hartree-Fock method and include the screening effects of the spatially structured dielectric environment semi-classically via an electrostatic picture as in Ref.~[\onlinecite{rosner_two-dimensional_2016}]. As a basis set for the electron-hole Hamiltonian, we use Slater determinants of dressed electron and hole wave functions which result from the Hartree-Fock calculations. Then, the many-body wave functions of the excited states are linear combinations of these determinants and we are left with a two-particle Schr{\"o}dinger equation.
	%left with the Bethe-Salpeter equation.\cite{sham_many-particle_1966} 
	A detailed description of the construction of the electron-hole Hamiltonian can be found Appendix \ref{appendix:hamiltonian}.
	After diagonalizing the electron-hole Hamiltonian, we obtain the many-body eigenenergies $E_\lambda$ and eigenstates $\ket{\Psi^\lambda}$. 
	In order to simulate an experimentally easily accessible property, we calculate the liner optical spectrum within the dipole approximation (for details see Appendix \ref{appendix:dipole}).

	The linear absorption spectrum is obtained by Fermi's golden rule:
    \begin{align}
        I(E) = \sum_{\lambda} \frac{2 \pi}{\hbar} \left| \braket{\Psi^\lambda|H_d | 0 } \right|^2 \delta (E_\lambda - E_0 - E). 
    \label{eq:lin_abs}
    \end{align}
    Here, $H_d$ is the light-matter coupling Hamiltonian, $E$ is the absorption/emission energy  and 
    $\ket{0}$ and $\ket{\Psi^\lambda}$ are the semiconductor vacuum state and many-body final state, respectively, with energies $E_0$ and $E_\lambda$,
    the former of which will be, as usual, set to zero in the remainder of this paper. %We consider unpolarized light here, see Supporting Information. 

	On similar grounds, we can calculate the total two-particle or "excitonic" density of states (DOS) %, which inhibits the whole electron-hole excitation spectrum
	\begin{align}
		A(E) = 2 \pi  \sum_\lambda  \delta (E -E_\lambda) \label{eq:two-part-dos},
	\end{align}
	which includes all possible two-particle excitations regardless of the selection rules, i.e., also accounts for "dark" excitons\cite{poem_accessing_2010,mullenbach_probing_2017,ye_probing_2014}.
	 To investigate excitonic correlations in real space, we use the spatially resolved two-particle DOS, which can be experimentally measured by using a dual-tip scanning tunneling microscope \cite{niu_double-tip_1995,okamoto_ultrahigh_2001,grube_stability_2001}:
	\begin{align}
		A(E, \textbf{r}_e, \textbf{r}_h ) = 2 \pi  \sum_\lambda \left| \Psi^\lambda \left( \textbf{r}_e, \textbf{r}_h \right)   \right|^2 \delta (E -E_\lambda). \label{eq:spectralFunction}
	\end{align}
	Here, $\vec{r}_{e/h}$ describe the position of the electron/hole. Further details on the calculation of the spatially resolved eigenstates $\Psi^\lambda \left( \textbf{r}_e, \textbf{r}_h \right)$ can be found in Appendix \ref{appendix:mb_eigenstates}.\footnote{For numerical reasons, the $\delta$-distribution was broadened using a Lorentzian with a full width at half maximum of $\Gamma = 0.01$\,eV throughout this work. }

\section{Tight-binding models for 2d semiconductors}
     In order to emulate 2d semiconductors, we use two different models. First, we consider a "monolayer" that consists of two hexagonal layers on top of each other (in the following, this model will be called "hexagonal bilayer") embedded in a dielectric environment. The coupling between the two layers mimics hybridization effects similar to the $d$-orbitals in transition metal dichalcogenides like \MoS \ as explained in Ref.~[\onlinecite{rosner_two-dimensional_2016}] and Appendix \ref{appendix:model}. In the tight-binding approach, we only consider an in-plane nearest neighbour hopping $t^{\vec{RR}'} =: t$ and
     an out-of-plane hopping $t^{\vec{RR}'} = t_\perp$, where $t^{\vec{RR}'}$ gives the energy associated with an electron hopping from state $\ket{\vec{R}'}$ to $\ket{\vec{R}}$. 
     
	An additional class of embedded monolayers is modeled using a honeycomb lattice with broken sublattice symmetry, which has been widely used to study 2d semiconductors such as hBN\cite{catellani_bulk_1987,galvani_excitons_2016,fugallo_exciton_2015} as well as graphene commensurately stacked with hBN\cite{giovannetti_substrate-induced_2007,sachs_adhesion_2011,bokdam_band_2014,slawinska_energy_2010,kharche_quasiparticle_2011}. This model leads to a massive Dirac equation in the low energy limit and is in the following referred to as "Honeycomb lattice". Again, we only consider in-plane nearest neighbor hopping $t$. 
	
	As one particular example representative for 2d semiconductors we consider \MoS. Therefore, the on-site energies and hopping matrix elements are chosen as in Ref.~[\onlinecite{rosner_two-dimensional_2016}] for the hexagonal bilayer ($t = 1.7$\,eV) and in Ref.~[\onlinecite{xiao_coupled_2012}] for the honeycomb lattice ($t = 1.1$\,eV). 
	
	The in-plane hopping $t$ determines how easily electrons can be localized. Thereby, smaller $t$ corresponds to smaller hopping probabilities and thus easier localization. In the following discussion (c.f. Fig.~\ref{fig:homogeneous_optical_total}), this case will be referred to as "localized" models in contrast to "delocalized" models with larger $t$.

%-------------------------------------------------------------------
%%%%%%%%%%%%%%%%%%%%%%%%%%%%%%%RESULTS%%%%%%%%%%%%%%%%%%%%%%%%%%%%%% 
%-------------------------------------------------------------------
%	\paragraph{Results}
	
%-------------------------------------------------------------------	
%%%%%%%%%%%%%%%%%%%%%%HOMOGENEOUS ENVIRONMENT%%%%%%%%%%%%%%%%%%%%%%
%-------------------------------------------------------------------
\section{Monolayer of a semiconductor embedded in homogeneous dielectric environment}

\begin{figure}[h]
		\includegraphics[width=0.8\columnwidth]{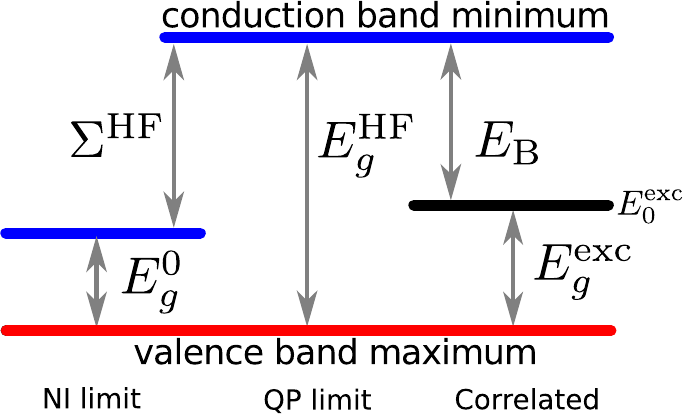}%
		\caption{\label{fig:band_scheme}(Color online) Schematic illustration of competing contributions to the excitonic gap $E_g^\t{exc}$:
			the single-particle band gap $E_g^0$, is widened to the dressed quasi-particle band gap $E_g^\t{HF}$ due to the exchange self-energy $\Sigma^\text{HF}$ which results from the ground-state electron-electron interaction. The optically measureable excitonic gap is reduced as compared to $E_g^\t{HF}$ due to the electron-hole binding energy $E_\t{B}$.}
	\end{figure}

\hfill \newline
	To discuss environmental influences on a monolayer, it is important to understand the interplay of excitonic binding and exchange self-energy effects. These effects are schematically shown in Fig.~\ref{fig:band_scheme}. In the non-interacting (NI) limit no Coulomb interaction is considered which results in the band gap being given by the single-particle band gap $E_g^0$. This single-particle gap is widened by the exchange self-energy $\Sigma^\text{HF}$ that enters the Hartree-Fock ground-state due to electron-electron interaction, resulting in the quasi-particle (QP) band gap $E_g^\text{HF}$ (QP limit). In the presence of Coulomb attraction between electron and hole, the excitonic gap $E_g^\t{exc}$ is strongly reduced by the exciton binding energy $E_\text{B}$ and cancels to some extent the exchange effects of $\Sigma^\text{HF}$. 
	
	\begin{figure}[th]
		\includegraphics[width=1\columnwidth]{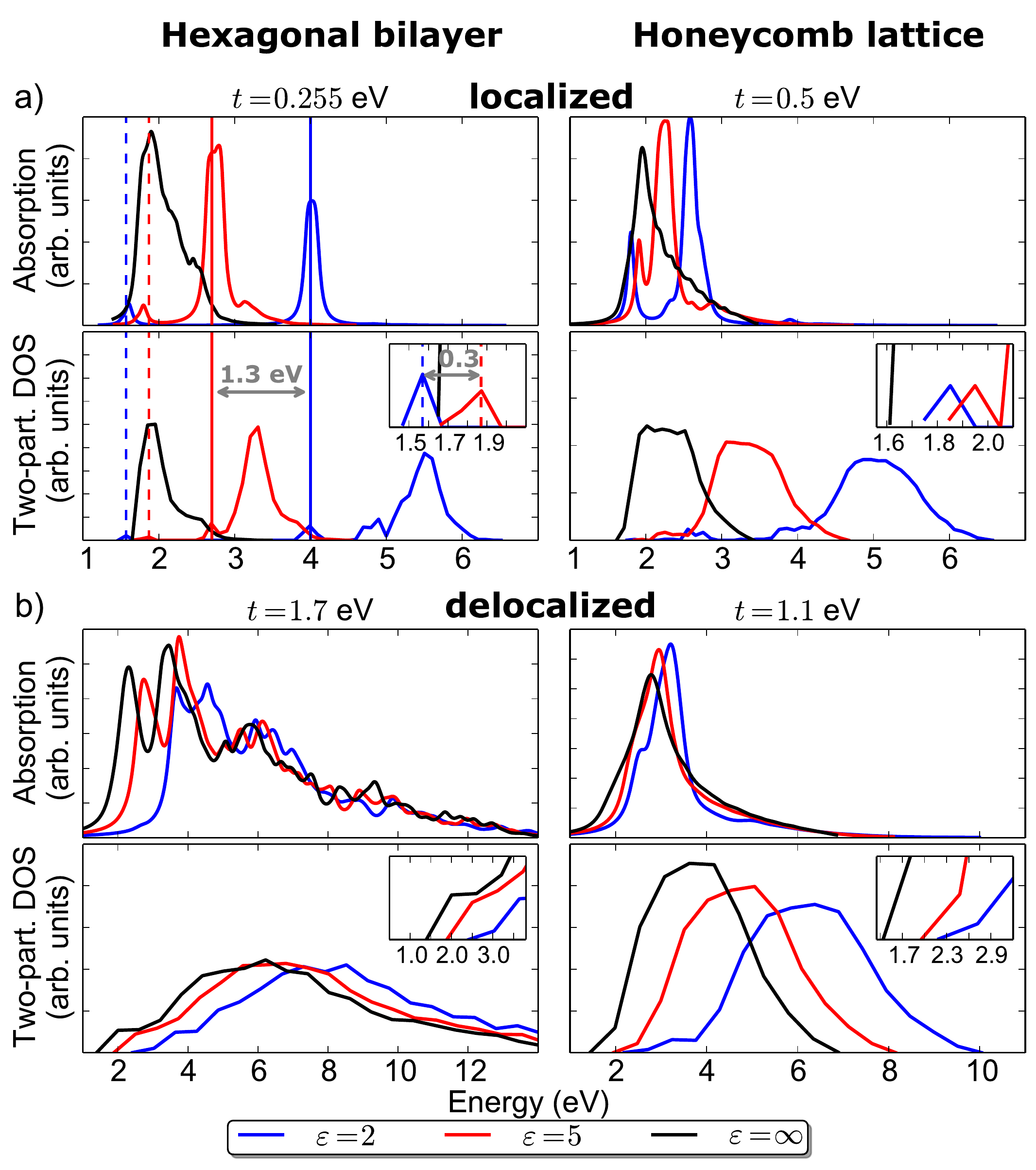}%
		\caption{\label{fig:homogeneous_optical_total}(Color online) Linear optical absorption spectra and total excitonic density of states for different dielectric constants of the environment of the monolayer for the localized (a) and delocalized (b) hexagonal bilayer (left panels) and honeycomb lattice (right panels). In the insets of the two-particle DOS the lowest-energy excitations are shown in more detail. High $\varepsilon$ correspond to small Coulomb interactions leading in our model systems to the description of independent particles for $\varepsilon = \infty$. %\MoS\ is emulated by the "delocalized" models.
			}
	\end{figure}

	To investigate the effect of external screening on excited states, we show two quantities in Fig. \ref{fig:homogeneous_optical_total} : (i) The linear optical absorption spectra obtained according to Eq. (\ref{eq:lin_abs}), and (ii) the two-particle DOS from Eq.~\eqref{eq:two-part-dos} for a real space supercell of approximate area $(9 \times 10)\,a^2$  using a lattice constant $a_\t{latt} = 3.18\,$\AA\ for \MoS \cite{komsa_effects_2012}.
		
	\begin{figure*}[t]
		\includegraphics[width=\textwidth]{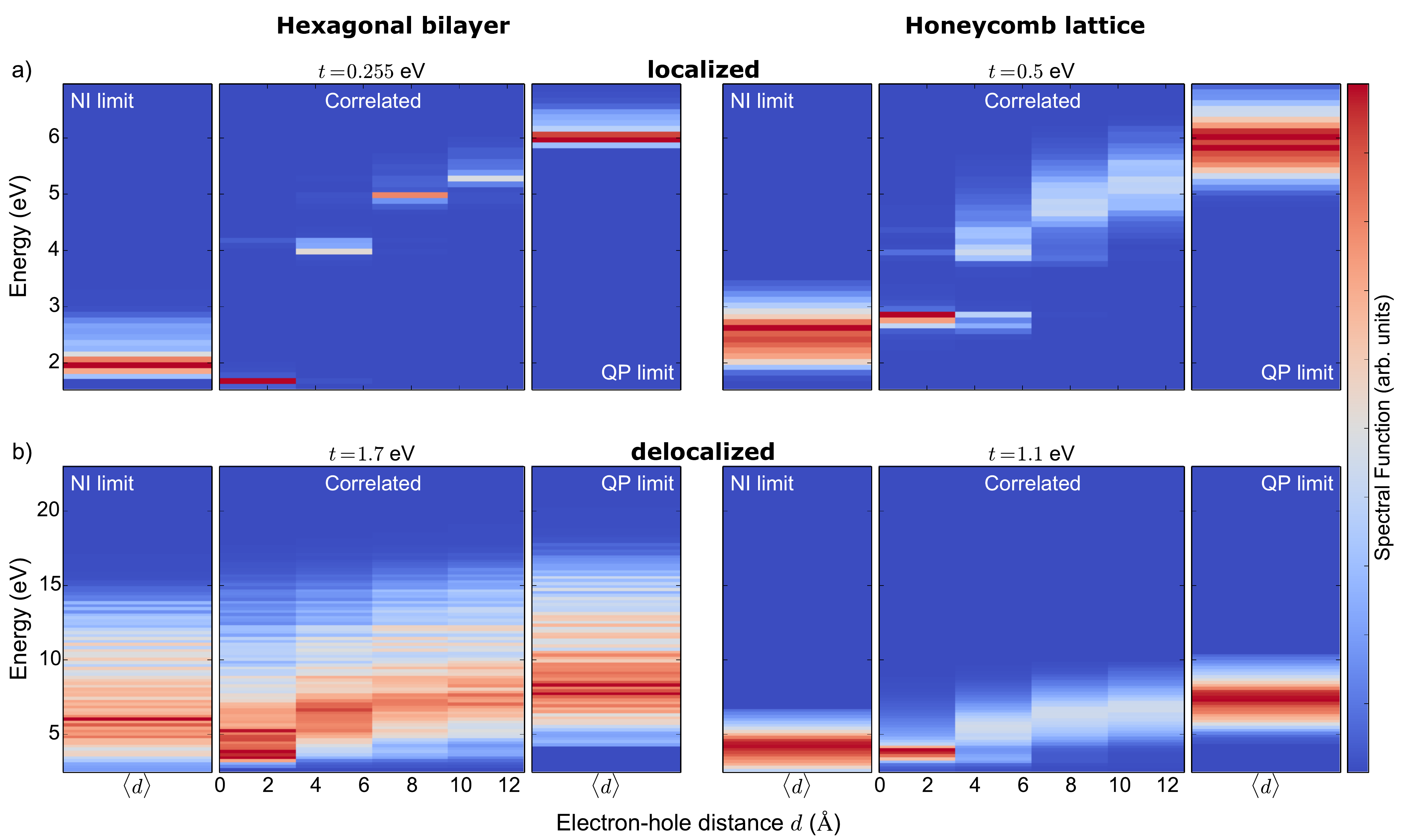}%
		\caption{\label{fig:SpectralFuncMap}(Color online) Local two-particle DOS for $\varepsilon = 2$ for different models and in-plane hopping parameters $t$. A smaller $t$ describes more strongly localized electrons. The quasi-particle (QP) limit corresponds to the case of vanishing electron-hole interaction ($V_{ijkl}^{ehhe} = 0$) but electron-electron interaction being included via the Hartree-Fock self-energy. In addition, the non-interacting (NI) limit $\varepsilon \rightarrow \infty$ is depicted. As the QP and the NI limit show no spatial dependence, the corresponding two-particle DOS are spatially averaged. Note that in the "localized" panels a broadening of $\Gamma = 0.01$\,eV and in the "delocalized" panels $\Gamma = 0.1$\,eV was used.
		 }
	\end{figure*}
	
	In general, for different models and in-plane hoppings $t$ we see  that external screening has a large impact on both the absorption spectra and two-particle DOS. In all cases, even though in the localized models most pronounced, smaller dielectric constants (i.e., weaker screening / stronger interaction) lead to energetically more extended spectra with spectral weight being shifted towards higher excitation energies. The NI limit is described by $\varepsilon \rightarrow \infty$ and thus yields all possible excitations of non-interacting particles. Then, the lowest-energy transition is the single-particle gap $E_g^0$. We see, that the influence of the environment is stronger on the the higher energy states, which are marked by solid lines in Fig.~\ref{fig:homogeneous_optical_total}a. They shift much more upon changes in the dielectric surroundings than the lowest-energy excitation (dashed lines), which are shown in the insets of the two-particle DOS. For example, in the localized hexagonal bilayer, the lowest exciton is shifted by $0.3\,$eV, which is $1$\,eV smaller than the shift of one of the higher energy excitations ($1.3\,$eV).

	To explain this relation between excitation spectra and dielectric environment, we show the two-particle DOS as function of electron-hole distance of the fully interacting system in Fig.~\ref{fig:SpectralFuncMap} for $\varepsilon = 2$. We compare this data to the non-interacting two-particle DOS and to the two-particle DOS in the QP limit to show the interplay of excitonic binding $E_\t{B}$ and exchange self-energy $\Sigma^\t{HF}$ as already discussed for Fig. \ref{fig:band_scheme}. In the NI and in QP limit, the figures show the joint DOS, i.e., the DOS of simple electron-hole excitations resulting from the non-interacting and Hartree-Fock calculations.  The lowest-energy excitation visible in these joint DOS are then the non-interacting $E_g^0$ and the HF quasi-particle band gap $E_g^\t{HF}$. As the NI and QP limits are always independent of the electron-hole interaction, there is no dependence of the spectral function on the electron-hole distance. % and we show a spatially averaged two-particle DOS for the non-interacting and the HF case, here. 
	
	As soon as the Coulomb interaction between the electron and the hole is considered (as shown in panels marked as "correlated") we find (i) a strong dependence on the electron-hole distance and (ii) a, in comparison to the QP band gap $E_g^\t{HF}$, reduced excitonic gap $E_g^\t{exc}$ which can be identified as the lowest peak in all correlated spectra. In more detail, we find that this lowest-excitation belongs to a bound exciton with the electron and hole beeing in close proximity. With growing electron-hole distance the correlated DOS shows excitonic peaks at elevated energies. Thus, the corresponding excitonic binding energies get smaller with increasing electron-hole distance until the correlated DOS approaches the QP limit.

	Consequently, higher energy states belong to two-particle excitations where electrons and holes are not in too close proximity. Since the environmental screening is more effective for larger separations of the electron and the hole, these higher states can generally be easily manipulated by engineering of the dielectric environment as recently also shown in Ref.~[\onlinecite{raja_coulomb_2017}].

\section{Response of lowest-energy exciton to dielectric screening}
\begin{figure}[ht]
		\includegraphics[width=\columnwidth]{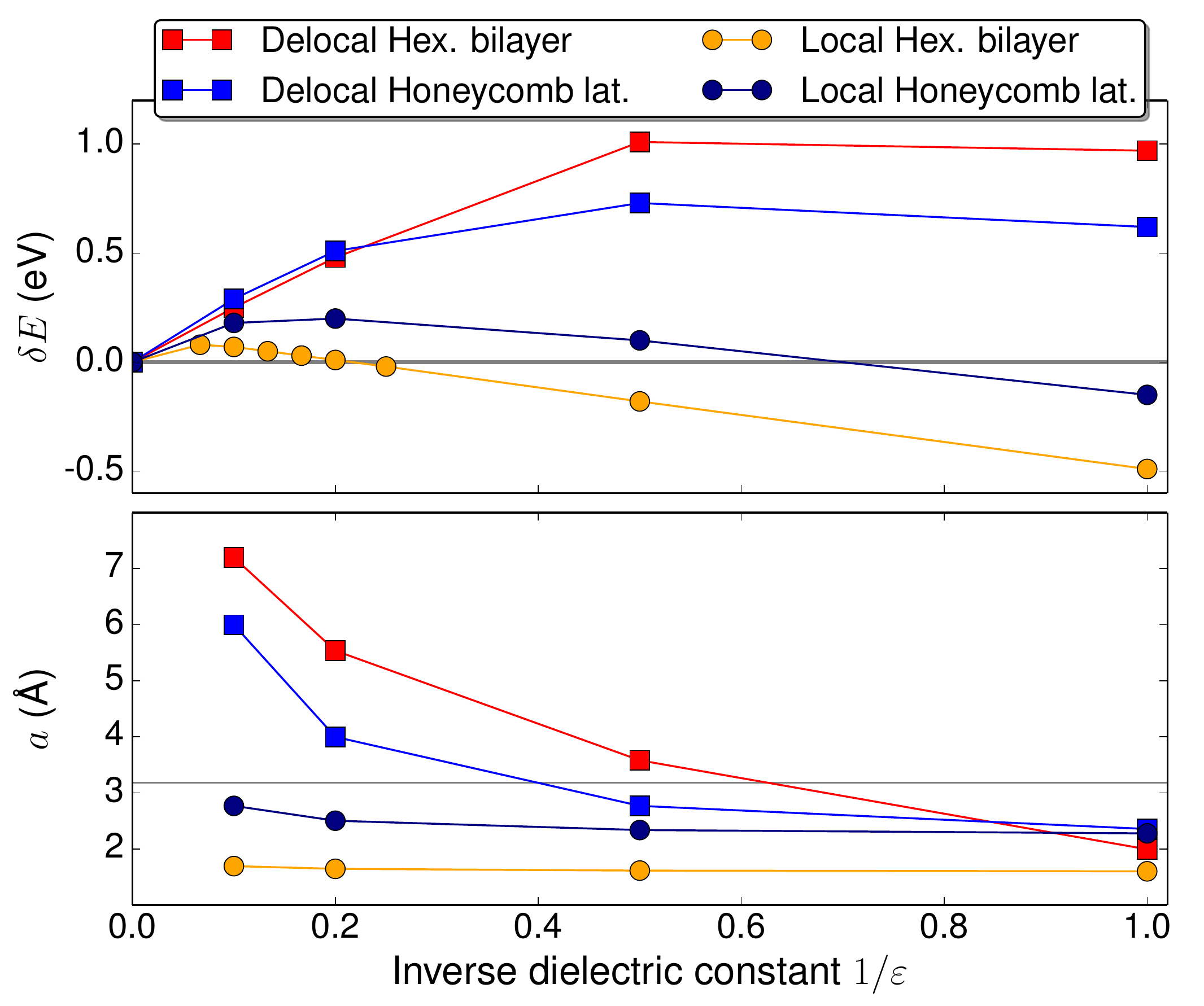}%
		\caption{(Color online) (Upper panel) Energetic shift \mbox{$\delta E=E_g^\t{exc}-E_g^0$} of the lowest-energy exciton $E_g^\t{exc}$ with respect to the non-interacting band gap $E_g^0$ as function of the inverse environmental dielectric constant $1/\varepsilon$. (Lower panel) The Bohr radius $a$ of the lowest-energy exciton as a function of $1/\varepsilon$. The gray line marks the lattice constant.}%
		\label{fig:GS_Exc}%
	\end{figure}

The lowest-energy excitation can show a different response to the dielectric screening than the higher excitations. Especially in the localized models, an almost perfect cancellation of the exciton binding and electron-electron exchange effects can occur. We analyze this behavior by plotting the energetic shift \mbox{$\delta E=E_g^\t{exc}-E_g^0$} of the lowest-energy exciton $E_g^\t{exc}$ with respect to the non-interacting band gap $E_g^0$ and the corresponding excitonic Bohr radius $a$ as a function of the inverse environmental dielectric constant in Fig. \ref{fig:GS_Exc}. In all cases, we see an increase of $\delta E \propto 1 / \varepsilon$ at sufficiently weak interactions ($1 / \varepsilon \ll 1$). As $E_g^\text{exc}$ and with it $\delta E$ depends on the size of the exchange self-energy $\Sigma^\t{HF}$ and the binding energy $E_B$ (c.f., Fig.~\ref{fig:band_scheme}), this directly follows from $\Sigma^\text{HF} \propto 1 / \varepsilon$ (for $1 / \varepsilon \ll 1$) and $E_B \propto 1 /  \varepsilon^2$ within the Wannier-Mott model for excitons\cite{czycholl}. In other words for sufficiently weak interactions ($1 / \varepsilon \ll 1$) exchange self-energy effects always dominate over electron binding effects, and the lowest-energy exciton shifts similarly upon changes in $1/\varepsilon$ as the higher excitons. However, the dependence of $\delta E$ on $1 / \varepsilon$ changes fundamentally at larger interaction strength being no longer given by a linear increase, but a decreasing energy shift upon increasing interaction. This can be seen exemplarily in the delocalized models for $1 / \varepsilon > 0.5$. The corresponding Bohr radii $a$ (see Fig.~\ref{fig:GS_Exc} lower panel, for calculation details see Appendix \ref{appendix:bohr_radius}) show that this change in the dependence of the excitonic gap on the Coulomb interaction occurs when $a$ approaches the length scale of the lattice spacings (marked as gray line). Then, the Wannier-Mott model is no longer a good description of the exciton binding. Thus, for the lowest-energy exciton a subtle interplay between exchange effects and the exciton Bohr radius determines how this exciton reacts on changes in environmental screening and it can either shift towards lower or higher energies upon increasing $1 / \varepsilon$. 

\section{Excitonic properties of a semiconductor monolayer embedded in laterally structured environment}
It has been suggested\cite{rosner_two-dimensional_2016} and experimentally realized\cite{raja_coulomb_2017} to create "junctions" with electronic band gap modulations within one and the same material by a laterally structured dielectric environment. Now, we investigate how two-particle excitations react to such structured dielectrics. 
		\begin{figure}[ht]
			\includegraphics[width=0.8\columnwidth]{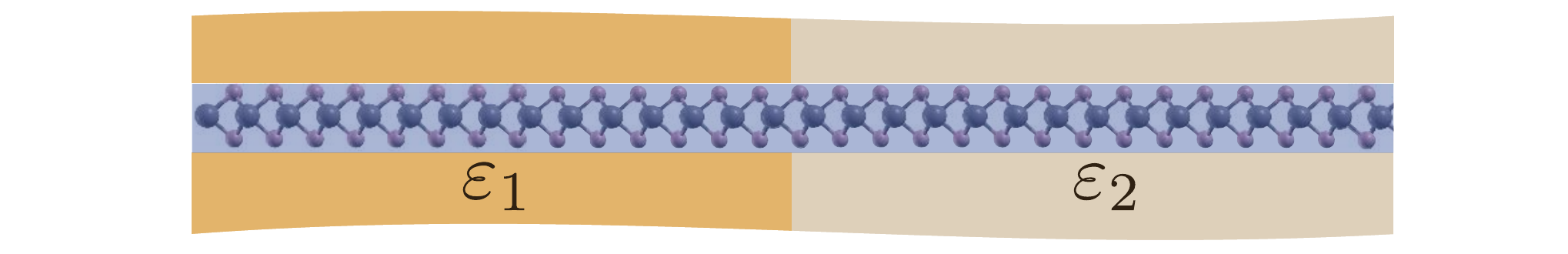} \\
			\includegraphics[width=0.8\columnwidth]{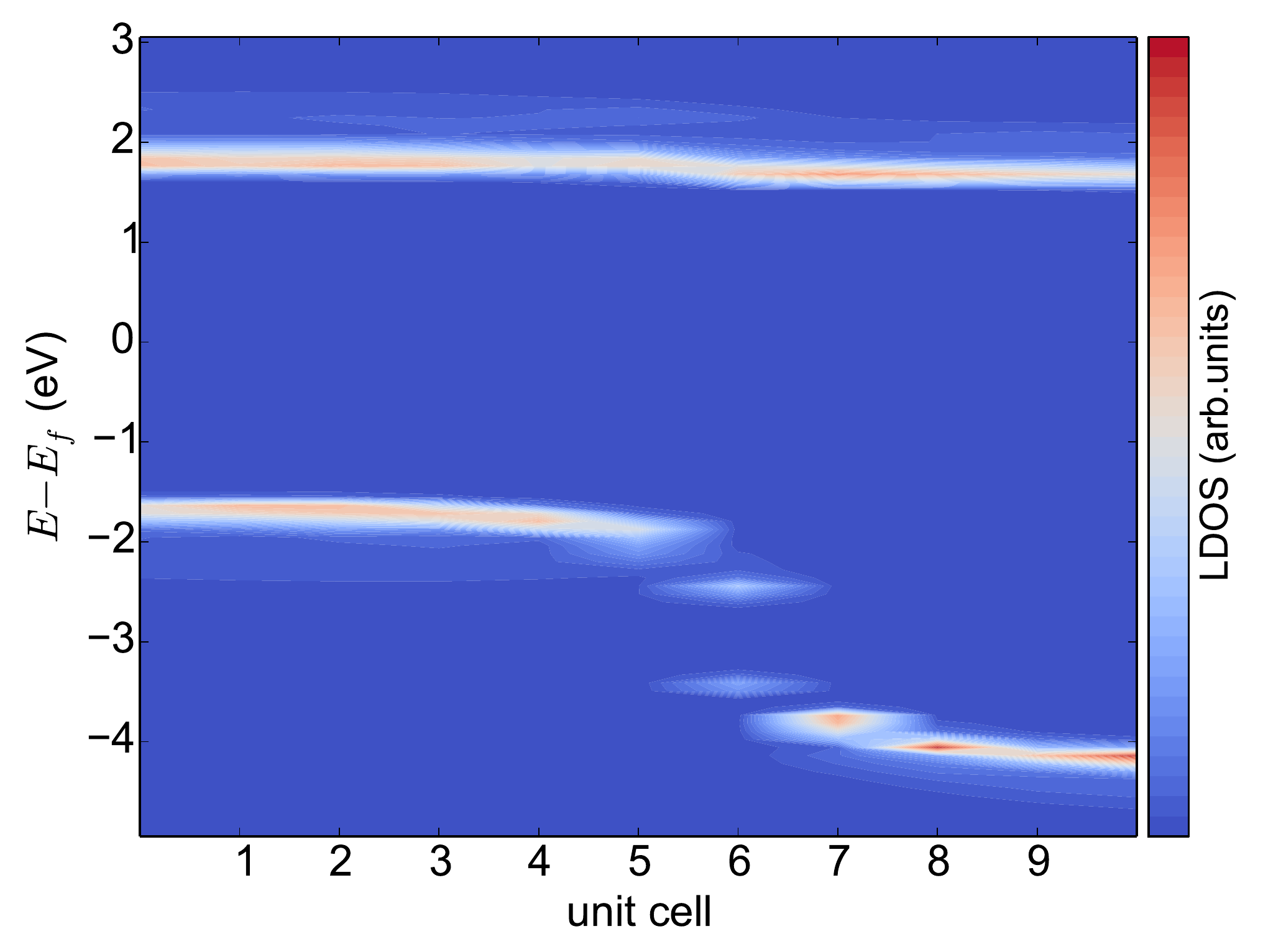}%
			\caption{\label{fig:one-part-ldos}(Color online) One-particle local DOS for different unit cells along a line perpendicular to the dielectric interface of the environment located in the unit cell 5 (left $\varepsilon_1 = 5$ and right $\varepsilon_2 = 2$). }
		\end{figure}
As the forthcoming results do not critically depend on the choice of the model Hamiltonian, we focus on the hexagonal bilayer with $t=0.255$\,eV. We use a laterally structured environment with two different dielectrics with $\varepsilon_1 = 5$ and $\varepsilon_2 = 2$. 

On the quasi-particle level this setup leads to a heterojunction-like spatial modulation of the band gap with a type-II like band line-up\cite{rosner_two-dimensional_2016}, as can be seen from the quasi-particle local DOS shown in Fig.~\ref{fig:one-part-ldos}: The dielectric interface induces a spatial dependence in the local DOS such that the quasi-particle band gap $E_g^\t{HF}$ in the region with the higher value of $\varepsilon$ is smaller than the band gap in the other region.
		
\begin{figure*}[ht]
			\includegraphics[width=2\columnwidth]{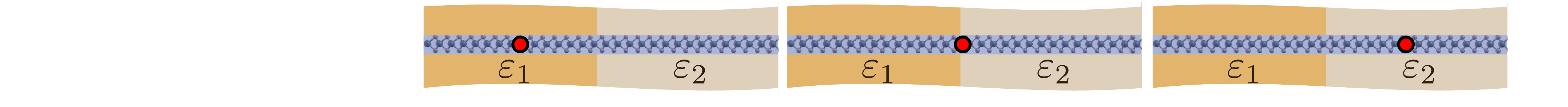}
			\includegraphics[width=2\columnwidth]{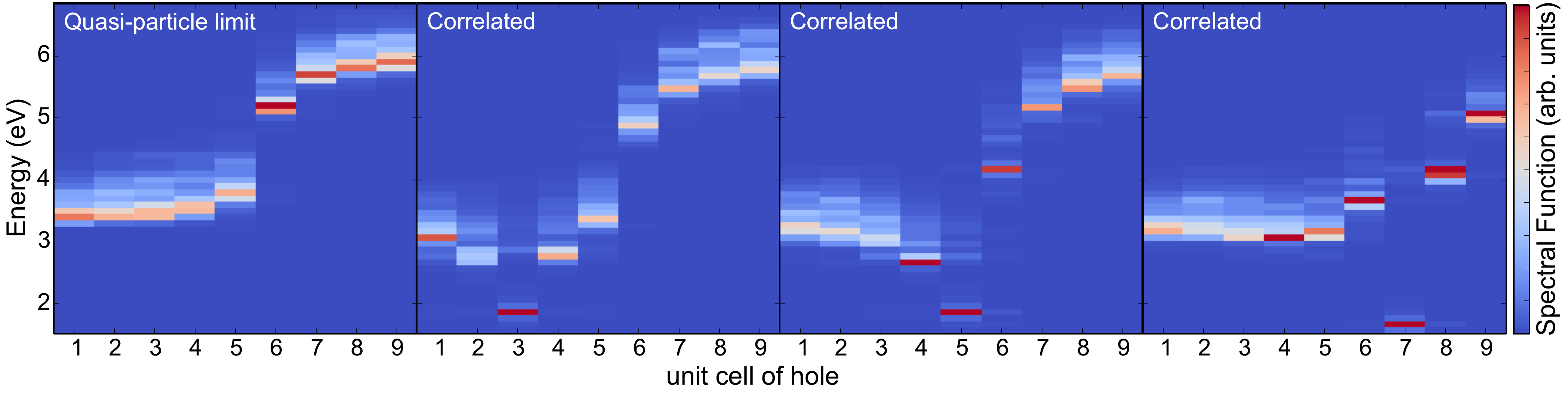}%
			\caption{\label{fig:ldos_heterogen}(Color online) Local two-particle DOS for a heterogeneous dielectric environment with $\varepsilon_1 = 5$ and $\varepsilon_2 = 2$. The $x$-axis marks the used hole positions whereas the red dot in the picture reveal the chosen electron positions for each subfigure. The quasi-particle limit (left panel) show all possible electron-hole excitations without electron-hole interaction. }
		\end{figure*}

The same physics governs the spatial energy landscape experienced by two-particle-excitations, as can be seen from the spatially resolved two-particle DOS shown in Fig.~\ref{fig:ldos_heterogen}. Here, red dots mark the position of the electron (to be more precise, the unit-cell index of the corresponding electron wave function). The unit cell index of the hole is used as a spatial coordinate across the supercell along a line perpendicular to the interface. In addition, we also show the spatially resolved quasi-particle limit of the two-particle DOS in the left panel which simply follows the heterojunction type band gap structure seen in the single-particle DOS of Fig.~\ref{fig:one-part-ldos}.\footnote{As the conduction band states are only weakly modified by the dielectric environment, the two-particle DOS in the quasi-particle limit is almost independent of the electron position and only the dependency on the hole position is shown in the left panel of Fig.~\ref{fig:ldos_heterogen}.}
	
For the interacting case, shown in the right three panels of Fig.~\ref{fig:ldos_heterogen}, we find that particularly the higher excitonic states are strongly influenced by the environment. Their energies follow the heterojunction-like band gap profile imprinted by the external dielectric essentially \textit{on an atomic scale}. This means that higher energy two-particle excitations can be trapped and guided on an atomic scale non-invasively by creating a spatial potential energy landscape as, for example, suggested in Fig.~\ref{fig:environment}e). 

The lowest-energy excitations, where electron and hole are closest together, are much less influenced by the dielectric environment, i.e., modulations in the corresponding excitation energies are about one order of magnitude smaller than for the higher energy excitons. Thus, while the higher excitations are trapped, the lowest-energy exciton can move almost freely.

Depending on the characteristics of the system, the energy landscape of the lowest exciton can either follow or be \textit{inverted} with respect to the energy landscape experienced by the higher energy excitons. Inverted in this context means, that the energies of the lower excitons shift in different directions as compared to their higher energy counterparts. The example considered in Fig.~\ref{fig:ldos_heterogen} reveals such an inverted potential energy landscape as the energies of the higher excitons in the region with $\varepsilon_1 = 5$ are smaller than in the region with $\varepsilon_2 = 2$ but the lowest exciton has a slightly larger energy in the region with $\varepsilon_1$. A comparison with Fig.~\ref{fig:GS_Exc} reveals that such inverted potential energy landscapes are expected for strongly localized excitons (i.e., Frenkel excitons) but not for Mott-Wannier excitons. Only for Frenkel excitons the lowest-energy excitation shifts towards smaller energies upon decreasing $\varepsilon$ (increasing $1 / \varepsilon$) in contrast to the shift of the higher energy excitations.

\section{Conclusions}
We have shown that two-particle excitations in 2d materials show a peculiar response to their dielectric environment. The higher energy states are in general strongly influenced by the dielectric environment, which means that their excitation energies can be tuned at energy scales comparable to the quasi-particle gap at spatial distances of a few lattice constants. This leads to the prospect of trapping and guiding higher energy excitons in 2d materials (including Rydberg excitons, which have been realized e.g., in \MoS\ \cite{hill_observation_2015}) simply via the dielectric landscape of the surrounding medium in a non-invasive way.

For the lowest-energy exciton, which determines by definition the excitonic band gap $E_g^\t{exc}$, we have shown that the Bohr radius in comparison to the lattice spacing determines how the excitonic gap responds to modulations in the dielectric environment. In the Wannier-Mott limit of delocalized excitons, the excitonic gap follows the trend of the higher energy excited states and, the quasi-particle gap, i.e., the ground-state exciton shifts in the same direction (albeit by a lesser amount) as the higher energy excitons upon changes in the dielectric environment. In the limit of strongly localized excitons, this trend is reversed and the ground-state exciton experiences a potential energy landscape which is inverted with respect to the higher excited states. 

Rather unexpected types of potential energy landscapes can emerge from this different response of two-particle excitations to external dielectrics, where higher energy states are trapped on atomic scale but lower energy states move almost freely or are even expelled from the trapping regions. This opens a large parameter space for dielectric quantum engineering of nanostructures hosting exotic physical states of matter with a variety of imaginable applications. One possible candidate are substrates with periodically modulated dielectric constants such as shown in Fig.~\ref{fig:environment}e, that would allow for a nanoscale separation of higher energy and ground-state excitons. Locally occupation-inverted regions should occur in these systems already upon weak optical pumping, which would be interesting in context of laser applications. % Of course, issues like carrier dynamics, light matter coupling, more complex many-body effects such as the formation of multi-excitons coupling in such systems are completely unexplored in the nanostructure discussed but define in our view a very promising ground for future exploration. 

\begin{acknowledgments}
This work was supported by the European Graphene Flagship, DFG via SPP 1459, GRK 2247 and \mbox{CZ 31/20-1}.  
We would like to thank Archana Raja for insightful discussions about experiments on partially covered 2d materials. \\
D.M. would like to acknowledge R. Binder and N.-H. Kwong for helpful discussions about the intricacies of the light- matter coupling. \\
M.R. would like to thank the Humboldt Foundation for support.
\end{acknowledgments}

%-------------------------------------------------------------------
%%%%%%%%%%%%%%%%%%%%%%APPENDIX%%%%%%%%%%%%%%%%%%%%%%
%-------------------------------------------------------------------

\appendix

\section{Electron-Hole Hamiltonian} \label{appendix:hamiltonian}
	The many-body wave functions of the excited state in the electron-hole picture can be described as linear combinations of Slater determinants of electron and hole wave functions. Upon single excitation of only one electron-hole pair, all electron-electron ($V_{ijkl}^{eeee}$) and hole-hole ($V^{hhhh}_{ijkl}$) interaction terms vanish. If we neglect electron-hole exchange-like terms (which do not contribute to the energy scale discussed in this paper), the many-body Hamiltonian in second quantization is given by
	\begin{align}
		H = \sum_i E_i^e e_i^\dagger e_i + \sum_i E_i^h h_i^\dagger h_i - \sum_{ijkl} V_{ijkl}^{ehhe} e_i^\dagger h_j^\dagger h_k e_l, \label{eq:twoPartHam}
	\end{align}
	where $E_i^{e/h}$ are dressed electron/hole eigenenergies from the Hartree-Fock calculations. The operator $(e/h)_i^{(\dagger)}$ annihilates (creates) an electron/hole in the Hartree-Fock eigenstate $\ket{\psi^i_{(e/h)}} = \sum_{\vec{R}} c_{\vec{R},{(e/h)}}^i \ket{\vec{R}}$. The Coulomb matrix elements $V_{ijkl}^{ehhe}$ between electrons and holes are then given by
	\begin{align}
		V_{ijkl}^{ehhe} = \sum_{\vec{R}_1, \vec{R}_2, \vec{R}_3, \vec{R}_4} c_{\vec{R}_1,e}^{i*} c_{\vec{R}_2,h}^{j*} c_{\vec{R}_3,h}^k c_{\vec{R}_4,e}^l \notag \\
		\times \bra{\vec{R}_1} \bra{\vec{R}_2} U(\vec{r} - \vec{r}') \ket{\vec{R}_3} \ket{\vec{R}_4}.
	\end{align}
	Here, $U$ describes the screened Coulomb interaction. Due to the orthogonality and the localization of the states $\ket{\vec{R}}$ we only consider two-center contributions, 
	\begin{align}
		V_{ijkl}^{ehhe} &\approx \sum_{\vec{R}, \vec{R}'} c_{\vec{R},e}^{i*} c_{\vec{R}',h}^{j*} c_{\vec{R}',h}^k c_{\vec{R},e}^l \notag \\
		&\times \bra{\vec{R}} \bra{\vec{R}'} U(\vec{r} - \vec{r}') \ket{\vec{R}'} \ket{\vec{R}}.
	\end{align}
	For further details see, e.g., Ref.~[\onlinecite{schulz_tight-binding_2006}] and the references therein. 
	In our lattice-discretized approach, the short-range spatial distribution of
	the states $\ket{\vec{R}}$ is not explicitly known. Thus, the electron-hole Coulomb matrix elements are approximated by 
	\begin{align}
		V_{ijkl}^{ehhe} = \sum_{\vec{R},\vec{R}'} c_{\vec{R},e}^{i*} c_{\vec{R}',h}^{j*} c_{\vec{R}',h}^k c_{\vec{R},e}^l U_{\vec{R},\vec{R}'}.
	\end{align}
	The Coulomb matrix elements $U_{\vec{R},\vec{R}'}$ contain the screening effects of the spatially structured dielectric environment:
% 	where we use an Ohno potential \cite{ohno_remarks_1964}
	\begin{align}
		U_{\vec{R}, \vec{R}'} = \frac{1}{\varepsilon_{\vec{R}, \vec{R}'}} \frac{e^2}{\sqrt{(\vec{R} - \vec{R}')^2 + \delta^2}}.
	\end{align}
	Here, $e$ is the elementary charge and the parameter $\delta$ takes into account the finite spread of the orbitals $\ket{\vec{R}}$ for $\vec{R} = \vec{R}'$. To emulate \MoS\ we choose  $\delta = 1.5$\,\AA\ according to Ref.~[\onlinecite{rosner_two-dimensional_2016}].
	%which leads to a bare on-site potential $U_{\vec{R}\vec{R}} = 9.6$\,eV. This corresponds to a density-density matrix element in \MoS\ for the $d_z^2$ orbitals \cite{rosner_two-dimensional_2016}.
	The macroscopic dielectric function $\varepsilon_{\vec{R}, \vec{R}'}$ includes the screening effects of the environment. For a homogeneous environment it is set to a constant value, whereas in a heterogeneous environment the interface is included by using image charges as described in detail in \mbox{Ref.~[\onlinecite{rosner_two-dimensional_2016}]}.

\begin{table*}[thp]
\centering
\begin{tabular}{cl|cccccc}
\hline
                                                       &         & $t$      & $t_{ii}^A$ & $t_{ii}^B$  &$a$  & $t_\perp$ & c   \\ \hline
\multicolumn{1}{c}{\multirow{2}{*}{Hexagonal bilayer}} & Localized   & $0.225\,$ eV & $0.0\,$ eV       & $0.0\,$ eV                         & $3.18\,$\AA & $-0.85\,$eV     & $a/4$ \\ 
\multicolumn{1}{c}{}                                   & Delocalized & $1.7\,$ eV   & $0.0\,$ eV       & $0.0\,$ eV                           & $3.18\,$\AA & $-0.85\,$eV    & $a/4$ \\ \hline
\multirow{2}{*}{Honeycomb lattice}                     & Localized   & $0.5\,$ eV      & $0.83\,$ eV   & $-0.83\,$ eV                        & $3.19\,$ \AA &    &     \\  
                                                       & Delocalized & $1.1\,$ eV      & $0.83\,$ eV   & $-0.83\,$ eV                          & $3.19\,$ \AA &      &     \\ \hline
\end{tabular}
\caption{Tight-binding model parameter for the description of MoS$_2$ \label{tab:Params}}
\end{table*}

	\section{Dipole matrix elements}  \label{appendix:dipole}

 	To directly investigate experimentally easily accessible optical properties, we calculate the
 	linear optical absorption spectrum. In the dipole approximation,
 	the light-matter coupling can be described by the dipole Hamiltonian: 
 	\begin{align}
	 	H_d = -e \vec{E} \sum_{ij} \underbrace{\braket{\psi^i_e | \vec{r} | \psi^j_h}}_{\textbf{d}_{ij}^{eh}} e_i h_j + h.c.
	 \end{align} 

 	Its matrix elements, that contain the optical selection rules
 	can unambiguously be obtained from matrix elements
 	of the position operator $\vec{r}$ between the quasi-particle electron and hole 
 	states $\ket{\psi^i_{e/h}}$,
 	\begin{align}\label{eq:dipoleME}
     \mathbf{d}^{eh}_{ij}=e\langle\psi^{i}_e|\vec{r}|\psi^{j}_h \rangle.
 	 \end{align}
 	 Here $e$ is the electron charge and $\vec{r}$ is (in consistency
 	 with the spatial resolution on a lattice scale) approximated by the 
     lattice operator $\vec{r} \approx \sum_{\vec{R} } 
     \ket{\vec{R}}\vec{R} \bra{\vec{R}}$,~\cite{leung_electron-hole_1997}
 	 thus neglecting the short range contributions that are not accessible in the
 	 tight-binding model.
 	 Depending on the polarization of the electric field vector of
 	 the incident light, different vectorial
 	 components are projected out of $ \mathbf{d}^{eh}_{ij}$ in Eq.~(1). 
 	 We simulate the case of unpolarized light via an equally weighted
 	 superposition of $I(\vec{E}=[100])$, $I(\vec{E}=[010])$ and $I(\vec{E} = [001])$.
 	 
	\section{Many-body eigenstates} \label{appendix:mb_eigenstates}
 
	For the spatially resolved two-particle DOS, Eq.~(3) in the main text, we calculate the eigenstates $\Psi^\lambda (\vec{r}_e, \vec{r}_h)$ of the electron-hole Hamiltonian (Eq.~\eqref{eq:twoPartHam}) as a function of the electron and hole position $\vec{r}_{(e/h)}$. Therefore, we calculate the eigenstate $\ket{\Psi^\lambda}$ from the many-body Hamiltonian \eqref{eq:twoPartHam} as linear combination of Slater determinants of electron and hole wave functions build from the Hartree-Fock eigenstates $\ket{\psi^i_{(e/h)}} = \sum_{\vec{R}} c_{\vec{R},{(e/h)}}^i \ket{\vec{R}}$:
	\begin{align}
		\ket{\Psi^\lambda} &= \sum_{nm} a_{nm}^{\lambda} \ket{\psi_{(e)}^n} \ket{\psi_{(h)}^m} \notag \\
		&= \sum_{nm} a_{nm}^{\lambda} \sum_{\vec{R},\vec{R}'} c_{\vec{R},{e}}^n c_{\vec{R}',{h}}^m \ket{\vec{R}} \ket{\vec{R}'}.
	\end{align}
	We determine the expansion coefficients $a_{nm}^{\lambda}$ from numerically diagonalizing the many-body Hamiltonian.
	The spatial representation of the eigenstate is then the projection on the electron and hole position $\ket{\vec{r}_{(e/h)}}$:
	\begin{align}
		\Psi^\lambda (\vec{r}_e, \vec{r}_h) &= \bra{\vec{r}_e} \braket{\vec{r}_h |\Psi^\lambda} \notag \\ 
		&= \sum_{nm} \sum_{\vec{R},\vec{R}'} a^\lambda_{nm} c_{\vec{R},e}^n c_{\vec{R}',h}^m  \delta_{\vec{r}_e,\vec{R}} \delta_{\vec{r}_h, \vec{R}'}
	\end{align}

\section{Semiconductor tight-binding models} \label{appendix:model}

\begin{figure}
		a)	
		%\begin{minipage}{0.49\columnwidth}
			\raisebox{\dimexpr-\totalheight+\ht\strutbox\relax}{\includegraphics[width=0.4\columnwidth]{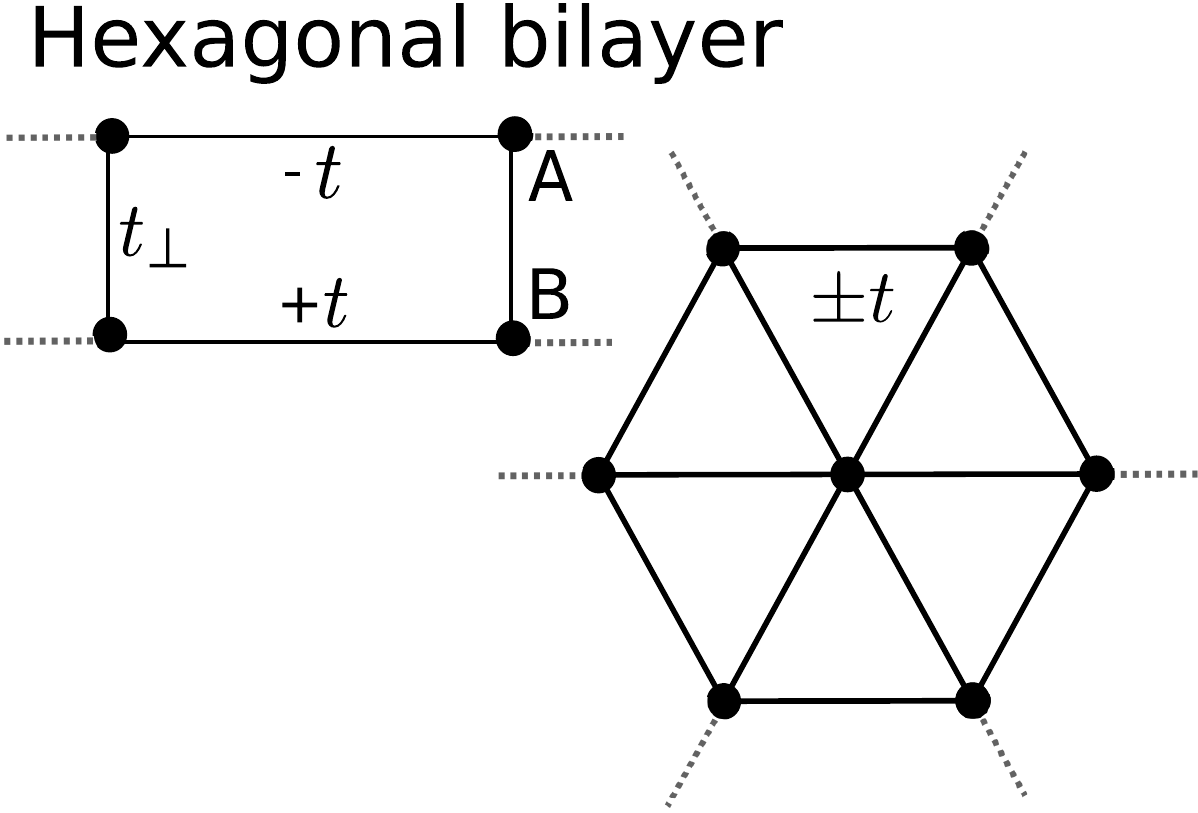} }
		%\end{minipage}
	\hfill b) 
		%\begin{minipage}{0.49\columnwidth}
			\raisebox{\dimexpr-\totalheight+\ht\strutbox\relax}{\includegraphics[width=0.4\columnwidth]{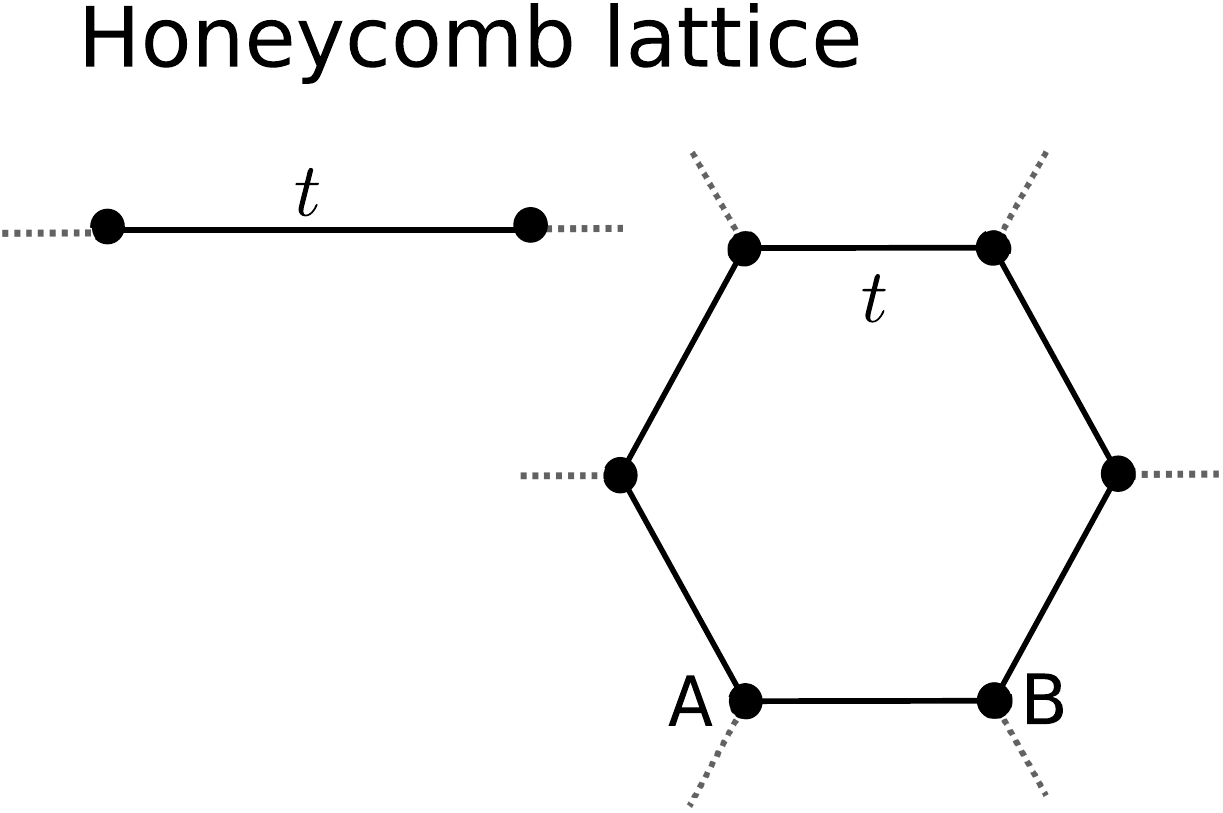}  }
		%\end{minipage}
	\caption{The 2d semiconductors are modeled using two different tight binding models labeled "hexagonal bilayer" and "honeycomb lattice", where the schematic side and top views of the model illustrate the hopping matrix elements. \label{fig:model}}
\end{figure}

From the perspective of tight-binding modelling, several methods exist to open a gap in an initially gapless electronic band structure. The two distinct models we use are depicted in Fig.~\ref{fig:model}, mimicing a hybridization gap (Fig.~\ref{fig:model}a) and a broken sublattice symmetry (Fig.~\ref{fig:model}b).

In a single \MoS\ layer, two sets of orbitals ($d_{z^2}$  and $\{d_{xy},d_{x^2-y^2}\}$) are responsible for the properties of the highest valence and lowest conduction band. The band gap results from hybridization between these blocks. To mimic these hybridization effects, we use a model which consists of two hexagonal layers placed in the distance $c$ on top of each other. Then the layer A and B mimic the $d_{z^2}$  and $\{d_{xy},d_{x^2-y^2}\}$ blocks, respectively. We only consider in-plane nearest neighbour hopping $t$ and out-of-plane hopping $t_\perp$. 

A different class of semiconducting monolayers can be modeled by using a honeycomb lattice with broken sublattice symmetry. This model leads to a massive Dirac equation in the low energy limit. Again, we only consider in-plane next nearest hopping $t$. 

For the hexagonal bilayer, the on-site energies and hopping matrix elements are chosen to reproduce DFT band gaps and band width of \MoS\ as in Ref.~[\onlinecite{rosner_two-dimensional_2016}]. For the honeycomb lattice the parameter from Ref.~[\onlinecite{xiao_coupled_2012}] are used. In both models we control how localized the electrons are with the in-plane hopping $t$. %Varying $t$ effectively tunes the spatial extent of the exchange hole surrounding all electrons. 
The hopping $t_{ii}^{A/B}$ defines the on-site energy and quantifies the sublattice symmetry breaking. The employed parameters and their notation are presented in Table~\ref{tab:Params}.

\section{Bohr radius} \label{appendix:bohr_radius}
	To analyze the spatial extent of the lowest-energy excitation $E_g^\text{exc} $, we calculate the corresponding excitonic Bohr radius $a$ for every dielectric constant $\varepsilon$. We use the expectation value of the Coulomb interaction $V$ between electron and hole:
	\begin{align}
		\braket{V_0} = \frac{e^2}{\varepsilon a}.
	\end{align}
	To calculate $\braket{V_0}$, we calculate the expectation value of the Coulomb matrix elements $V_{ijkl}^{ehhe}$ with the eigenstate which corresponds to $E_g^\text{exc}$:
	\begin{align}
		\braket{V_0} = \Bra{\Psi^0}  V_{ijkl}^{ehhe}  \ket{\Psi^0}
	\end{align}
	separately for each $\varepsilon$ of the homogeneous dielectric environment. 
 	 
\bibliography{references}
\end{document}